\documentclass[twocolumn]{aastex6}
\usepackage{multirow}

\collaborationName{Optics and Instrumentation Deparment}
\begin{document}

\title{Implementation of MUAV as reference source for GLAO systems}

\author{Ra\'ul Rodr\'iguez Garc\'ia\altaffilmark{1,2} and Salvador Cuevas \altaffilmark{1}}
\affil{Instituto de Astronom\'ia\\
Universidad Nacional Autonoma de M\'exico, Apdo. Postal 70-264, 04510 \\
 Ciudad de M\'exico, M\'exico}

\altaffiltext{1}{Instrumentation Department IA-UNAM}
\altaffiltext{2}{rrodriguez@astro.unam.mx}

\begin{abstract}

We propose an alternative method to generate an artificial reference source for a Ground Layer Adaptive Optics system airborne on a Multirotor Unmanned Aerial Vehicle. Turbulence profiles from the Ground Layer at the National Astronomical Observatory in San Pedro Martir were analyzed to establish the requirements in luminosity, altitude, and flight stability of an artificial source. We found the source must be at least 800 m above the observatory's surface and follow a fixed trajectory to emulate the apparent movement of the stars with a stability of 1.54 cm in time intervals smaller than 18 ms. We establish some commercial and customized MUAVs can nearly accomplish this task.
\end{abstract}

\keywords{Adaptive Optics, GLAO, Artificial Reference Source, MUAV, UAV.}

\section{Introduction} \label{sec:intro}

Modern astronomical instrumentation has focused big efforts on the development and construction of instruments for the new generations of terrestrial mega telescopes. Most will have to use Adaptive Optics (AO) systems, to correct the negative effects produced by the terrestrial atmosphere on the incident light, and thus obtaining its maximum performance. This technique has more than 60 years of development since it was proposed by  \citet{Babcock1953}. However, to apply it, we need luminous references to measure, in real time, the distortions of the wave-fronts coming from astronomical objects that reach the Earth  \citep{Hardy1998}.

The AO systems work with two types of reference sources: one uses natural guide stars (NGS). Nevertheless, these are insufficient to cover the entire observable sky. The other type emerges as a solution to the problem presented by the NGS and consists of implementing laser techniques (Laser Guide Star, LGS) \citep{Foy1985}.

Two methods are normally used: the LGS Sodium and the LGS Rayleigh. The first method takes advantage of the resonance of the atmospheric sodium layer produced by a laser (589.6 nm wavelength), which is emitted from the Earth's surface to a height of 90 km (average height of the sodium layer) to form a luminous spot, whose backscattering light returns through all the atmospheric layers until it reaches the telescope's optics. The second method uses the Rayleigh scattering produced by the air molecules in the atmosphere. However, since the air density decreases with height, this type of lasers only have a good performance to produce a bright spot in heights less than 20 km. 

It should be noted that even though the LGS allows to have luminous references in the whole observable field by terrestrial telescopes, its implementation is not so simple, and it needs specific special facilities and expertise, no to mention the cost \citep{Wizinowich2012}. 

According to the atmospheric turbulence characteristics, it is well known that the greatest distortion of the wave-front is mainly produced in two layers known as the Free Atmosphere, and the Ground Layer \citep{Skidmore2009}. The latter is where usually the greatest distortion of the wave-front occurs \citep{Tokovinin(2003), Egner(2007), Osborn(2010)}. Therefore, specialized AO systems have been developed to correct the effects of this layer; they are known as Ground Layer Adaptive Optics (GLAO) \citep{Tokovinin2010}.

Nowadays, there are new greater vertical resolution instruments that have come to improve the characterization of the Ground Layer, and now, we know that the greatest influence over incident wave-fronts is inside the first 100 m above the surface of some observatories \citep{Oya2015,Hickson2010,Tokovinin2010B}. In Mexico, at National Astronomical Observatory (OAN) some measurements with one of these instruments were performed, which we can notice a similar Ground Layer behavior \citep{Leonardo2015}. However, it is necessary to conduct a long term characterization campaign for confirming this scenario.

\pagebreak

Additionally, the development and application of Unmanned Aerial Vehicles (UAV) have been an international boom in different fields of science, including in astronomical instrumentation for optical telescopes such as is found in  \cite{Biondi2016}.

In \cite{Basden2018} a general and extensive analysis of using rotary unmanned aerial vehicles as an artificial guide star can be found. Here, it was proposed and analyzed the idea of a future UAV application, where the devices are used as an auxiliary reference source for OA systems. In this article, the device performance is simulated working together with laser references for the correction of several layers of atmospheric turbulence. Their results show the necessity to reach a height of 10 km, which is not at present possible. 

A key point for use MUAV in astronomical instrumentation applications is to determine the optical effects generated by these devices when flying. Rodr\'iguez Garc\'ia et al. (2019) measured these effects and determined that in isothermal conditions, the optical turbulence produced by these devices is negligible \citep{Raul2019}.

In this article, we analyze the possibility of the current implementation of an artificial reference source, when we restrict the turbulence correction to the Ground Layer. We compared the advances achieved by the technology with  Multirotor Unmanned Aerial Vehicle (MUAV) and the atmospheric conditions found in the observatory of San Pedro Martir in Mexico as an example for its application. 

We study the feasibility of using a MUAV (including programmable-ready to fly devices), as a carrier of a LED light source to generate an artificial reference source for AO. This reference source will use to determine the distortion on the wave-fronts produced by the Ground Layer turbulence, and thus implement it for terrestrial telescopes with GLAO systems. In Section 2, we specify the general conditions for reference sources of an adaptive optics system, and the characteristics required by a UAV. In section 3, we establish the restrictions for this source based on the characteristics of the local turbulence of the Ground Layer at San Pedro Martir observatory. Finally, in Section 4 we introduce the features for the selection of a MUAV, and we analyze their performance, both for commercial and for custom design devices. This analysis seeks to determine if these devices can fully comply with the before established restrictions.

\section{General Conditions} \label{Condiciones}

\subsection{Reference Source of an AO System}

In general, any reference source used for adaptive optics systems must fulfill the following conditions:
\vfill

\begin{itemize}
\item The light source will be punctual and bright enough to be detected by the wave-front sensor of the adaptive optics system.

\item The reference source will remain within the isoplanatic angle of the observed field during the time required for observations \citep{Fried(1982)}

\item The source will remain spatially stable inside the AO wave-front sensor.

\item This reference source will be located at a certain distance from the telescope along to the optical axis, in which the error caused by the cone effect is the minimum \citep{Fried(1994)}.
\end{itemize}

It is important to point out that LGS need to be stabilized with some NGS in the field of view (FoV) of telescope \citep{Tokovinin2016}. 

\subsection{Unmanned Aerial Vehicles (UAV)}

To have a functional system that can be optimally implemented in astronomical observatories, concerning UAVs in general, it is necessary to establish the following operating conditions:

\begin{enumerate}
\item The device will have the capability to transport and supply the energy for the reference source.

\item It will operate semi-automatically and remotely from a Ground Station.

\item It will perform the following functions: take off; reach a specific operation position and height correspondent to the pointing position of the telescope; return and land at the initial take-off position (Home).

\item It will be stabilized within the field of the isoplanatic angle of the telescope.

\item The device will have the possibility to receive and perform commands to change their position constantly, and follow a specific trajectory as well, corresponding to the telescope tracking.

\item It will have adequate security systems to avoid accidents that could put people at risk, either at the infrastructure of the observatory or for the device itself \cite{Basden2018}.
\end{enumerate}

\section{RESTRICTIONS FOR THE REFERENCE SOURCE}

In this section, we establish the restrictions for the reference source (minimum height, size of the light source, brightness, and speed of movement) from the Ground Layer atmospheric parameters (seeing $\epsilon_{0GL}$, Fried parameter $r_{0GL}$, isoplanatic angle $\theta_{0GL}$, and coherence time $\tau_{0GL}$) for the OAN in Mexico.

To establish the restrictions of the reference source, need to know the characteristics of the Ground Layer of the site where we are trying to correct by AO. This latter is achieved by calculating the atmospheric parameters from the profile of turbulence ($C_{n}^{2}$ profile data), obtained experimentally at each astronomical site.

We determined the atmospheric parameters by two procedures: in the first, we employed the average value of atmospheric seeing $\epsilon_{0GL}= 0.56$ [arsec] reported by \cite{Skidmore2009} and \cite{Avila2011} for the Ground Layer of the OAN with the equations of the atmospheric parameters \citep{Tyson2011}. The results for two wavelengths, $\lambda=500\:nm$ and $\lambda=1500\:nm$ (in brackets), are in table \ref{tabla:PSPM} as \textit{mean}.

The second was done using the data of OAN acquired with the latest generation Low-Layer Scidar (LOLAS-2) instrument \citep{Lolas-2}, the same equations of atmospheric parameters were used and are in table \ref{tabla:PSPM} as \textit{LOLAS-2}. For both analysis were considered a maximum height of the turbulent layer of $h=60 \, m$, and an average wind speed for the first kilometer of height above the surface of the observatory of $v = 2.3 \, m/s$, reported by \cite{Avila2003}.

\begin{deluxetable}{ccCrlc}
\tablecaption{OAN's Ground Layer atmospheric parameters.\label{tabla:PSPM}}
\tablecolumns{5}
\tablenum{1}
\tablewidth{0pt}
\tablehead{
\colhead{} & \colhead{Seeing} & \colhead{Fried} & \colhead{Isoplanatic} &
\colhead{Coherence} \\
\colhead{Parameter} & \colhead{} & \colhead{parameter} & \colhead{angle} &
\colhead{time} \\
\colhead{} & \colhead{$\epsilon_{GL}$} & \colhead{$r_{0GL}$} & \colhead{$\theta_{0GL}$} &
\colhead{$\tau_{0GL}$} \\
\colhead{} & \colhead{(arcsec)} &
\colhead{(cm)} & \colhead{(arcmin)} & \colhead{(ms)}
}
\startdata
Mean & $0.56$ [$0.45$] & $17.8\;$ [$66.6$] & $3.22$ [$12.0$]  & $24.4$  [$91.2$] \\  
LOLAS-2  & $0.73$ [$0.59$]& $13.7\;$ [$51.3$] & $2.57$ [$9.60$]& $18.7$ [$70.1$] \\  
\enddata
\tablenotetext{}{Values ​​in brackets calculated for $\lambda=1500\:nm$.}
\end{deluxetable}

In addition to the atmospheric parameters calculated above, there is another fundamental parameter for our analysis. This parameter is known as focal anisoplanatism or \textit{cone effect}, in which are related the diameter of the telescope ($d_{0}$), the height of the reference source ($H_{ref}$) and features of the turbulence layer (Ground Layer) evaluated as isoplanatic angle ($\theta_{0GL}$) , as shown in the following relation \cite{Tyler1994}.

\begin{equation}
d_{0}=2.88H_{ref}\theta_{0GL}
\label{coneeffect} 
\end{equation}

We used the data from LOLAS-2 to determine the restrictions for the proposed reference source since we considered these data as the worst condition values for the characteristics of the analyzed turbulence.

Table \ref{tabla:d0} shows the required height for our reference source, for two wavelengths, considering the cone effect for a telescope of 2.1 m and 6.5 m, current and future infrastructures of the OAN respectively. It is important to note that as the turbulent layer is very close to the surface of the observatory, the values of the isoplanatic angle obtained are in order of arc minutes. This latter allows us when we evaluate only the cone effect of the Ground Layer, to have telescope's apertures of considerable diameter with reference sources at low height, compared to the laser techniques LGS of conventional adaptive optics.

\begin{table}
\tablenum{2}
\centering
\caption{Height of reference source.} \label{tabla:d0}
\begin{tabular}{ccc}
\tablewidth{0pt}
\hline
\hline
Telescope Aperture & $\lambda$=500 nm & $\lambda$=1500 nm\\
\hline
\decimals
  2.1 m & 967 & 259   \\  
 6.5 m & 2992 & \textbf{800}  \\
\hline
\multicolumn{3}{c}{NOTE. - Height in meters above the observatory surface.}
\end{tabular}
\end{table}

The maximum angular size of a reference source for an adaptive optics system can be calculated using the Fried parameter $r_{0GL}$ shown in equation \ref{E-Ldimm}. With this value, we can obtain the linear dimension of our source by using the distance from the telescope to the source (800 m) to get \textbf{2.3 mm}.

\begin{equation}
\alpha_{max}=\frac{\lambda}{r_{0GL}}
\label{E-Ldimm} 
\end{equation}

Furthermore, by the photometric data of a standard LED, with 2000 mcd of luminous intensity, $30^{\circ}$ of viewing angle and  a distance of 800 m (height of reference source) we obtain an illuminance (E) of 3.125x$10^{-6}$ lx. To obtain the apparent magnitude ($m_{v}$), we use the next equation, with a value of 2.54x$10^{-6}$ corresponding to the illuminance of a star with $m_{v}=0$ \citep{Allen1973}.

\begin{equation}
m_{v}=-2.5log_{10}\frac{E}{2.54\times10^{-6}}
\label{E-Magnitud} 
\end{equation}

The result is a reference source of $\mathbf{m_{v}=-0.23}$ of apparent magnitude. It is important to mention that, because it is an LED source, it is possible to regulate its intensity through electronic devices, which could have sources of lesser magnitude.

Besides, it is necessary to counteract the apparent motion of the sky. This motion is due to the Earth rotation, and it is possible to prove that the maximum velocity of our reference source will be at the Zenith. In this way, we use the angular velocity of the Earth (15 arcsec/s) and the distance from the telescope to the MUAV (800 m) to get a velocity of\textbf{ 5.83 cm/s}.

Another quite important parameter to know is the stability of our reference source. The stability can be obtained by the relation between the field of view of the wavefront sensor (typically 4 arcsec), and the distance of the reference source from the telescope. For this parameter, we obtained a result of \textbf{1.54 cm}.

Table \ref{tabla:fuente} shows a summary of the results mentioned above for each restriction.

\begin{deluxetable}{ccccc}[h!]
\tablecaption{Reference source's requirements.\label{tabla:fuente}}
\tablecolumns{5}
\tablenum{3}
\tablewidth{0pt}
\tablehead{
\colhead{Minimum} &
\colhead{Linear} &
\colhead{Apparent} &
\colhead{Maximum} &
\colhead{Required} \\
\colhead{high} &
\colhead{dimension} & \colhead{brightness} & \colhead{speed} & \colhead{stability}\\
\colhead{(m)} &
\colhead{(mm)} & \colhead{($m_{v}$)} & \colhead{(cm/s)} & \colhead{(cm)}
}
\startdata
  800  & 2.3 & -0.23* & 5.83 & \textbf{1.54} \\  
\enddata
\tablenotetext{*}{ Value computed for standard LED.}
\end{deluxetable}


From the previous table the value of the stability of the UAV is highlighted, since this is considered the most restrictive parameter of the analysis, because in a theoretical case, the aerial device should keep the light source within a diameter of 1.54 cm (wavefront sensor FoV) at 800 m in height in a time interval of at least 18 ms (value of time, see table \ref{tabla:PSPM}.  In another way, due to the AO systems' performance is always better than the design features (obtained from the atmospheric parameters), an instrumental view could be to use the LGS fast steering mirror for the reference source image correction. Therefore, the requirement with allowing a little drift in the WFS FoV (1\%), and measuring at 1 kHz or each 1 ms, would result in 1.54/100 cm per ms or 15.4 cm per second.

\section{ Multirotor Unmanned Aerial Vehicles} \label{sec:MUAV}

For this project, we based the use of MUAV on the following features:

\begin{itemize}
\item Vertical take-off and landing (VTOL).
\item Flight static (hover) and trajectory tracking (lift at low horizontal and vertical speed).
\item Stability and resistance to disturbances (wind gusts).
\item Ease to change direction (maneuverability).
\item Ability to transport payload.
\item Currently, existing fully programmable options ready to fly.
\end{itemize}

\subsection{MUAV Performance}

The performance of a MUAV relates to the configuration of each device, which determines characteristics such as maximum flight height, top speed, time of autonomy, maximum capacity payload, stability, maneuverability, among others. Within these configurations, we can find elements as the number and type of motors and propellers, which sets up the maximum thrust of the propulsion systems; the flight controller that includes sensors, electronics, and control algorithms;  the frame, which gives mechanical support for all systems, and one critical factor is the battery used, which establishes the autonomy time of the devices \citep{Shi2017}.

There are a wide variety of applications and developments for MUAVs. Both for those of a commercial nature those of custom design as well. However, these last ones do not normally report complete data on the characteristics of their capabilities, beyond those needed for their use. Unlike the custom devices, commercial MUAVs such as those distributed by the company DJI provide the specifications and characteristics of most of its models on its website (www.dji.com).

Table \ref{tabla:Capmax} reports the values of the maximum operating capabilities of  MUAVs, obtained through the comparison of the information collected, both in the scientific literature and in that found on the DJI's website for each restriction mentioned above. In this table, we can see that MUAV are capable to reach a height of at least 2000 m above the OAN surface (2800 AMSL). The vertical and horizontal speeds are enough for the displacement of MUAV because they are several orders of magnitude above the requirement of maximum speed (see, Table \ref{tabla:fuente}). Furthermore, with the vertical velocity of 5 m/s, the MUAV takes \textbf{5.33 min} to ascend and descend to the operational height (800 m). Taking a maximum time of autonomy of 40 min, a positioning time of 1-5 min (to place the reference source in the FOV of the WFS and close the control loop), and a margin for the safety of 5 min, it would have around \textit{24 min} to operate as a reference source (the performance also depends on the type of battery and the environmental conditions). The effective observation time could be increased by in-flight synchronized swapping of MUAVs with a fresh battery \cite{Basden2018}. 

It is necessary to keep in mind that the maximum wind speed for the operation of the telescopes, at the OAN and other sites, ranges from $v_{max} = 15-20 \,\: m/s$.

\begin{deluxetable}{cccccc}
\tablecaption{MUAV maximum performance. \label{tabla:Capmax}}
\tablecolumns{3}
\tablenum{4}
\tablewidth{0pt}
\tablehead{
\colhead{Altitude} &
\colhead{Speed} &
\colhead{Wind} &
\colhead{Payload} &
\colhead{Autonomy} \\
\colhead{} &
\colhead{} &
\colhead{resistance} &
\colhead{} &
\colhead{} \\
\colhead{(AMSL)$^{1}$} &
\colhead{(m/s)} & \colhead{(m/s)} & \colhead{(kg)} & \colhead{(min)}
}
\startdata
\multirow{2}{*}{5000-6000} & H: 20 & \multirow{2}{*}{12} & \multirow{2}{*}{1 - 6} & \multirow{2}{*}{20 - 40} \\
                         & V: 5    &    &     \\        
\enddata
\tablenotetext{1}{ Above Mean Sea Level; H:Horizontal; V:Vertical.}

\end{deluxetable}%


\subsection{Positioning, Trajectory Tracking and Stability }

In this subsection, we determined if a MUAV can remain fixed, to the apparent movement of the astronomical objects during the observation time. For this condition, it is important to understand that this motion is due to the terrestrial rotation.  In other words, there is a point aligned with the rotation axis of the Earth that remains in a stationary position (near to Polaris Star), while objects close to this point appear to have a smaller movement compared to the rest of the sky. Conversely, objects that are $90^{\circ}$ from this point move with greater velocity and therefore, the MUAV will have to reach the maximum calculated travel speed (see, Table \ref{tabla:fuente}).

\begin{figure}[h!]
\center
\figurenum{1}
\includegraphics[scale=0.4]{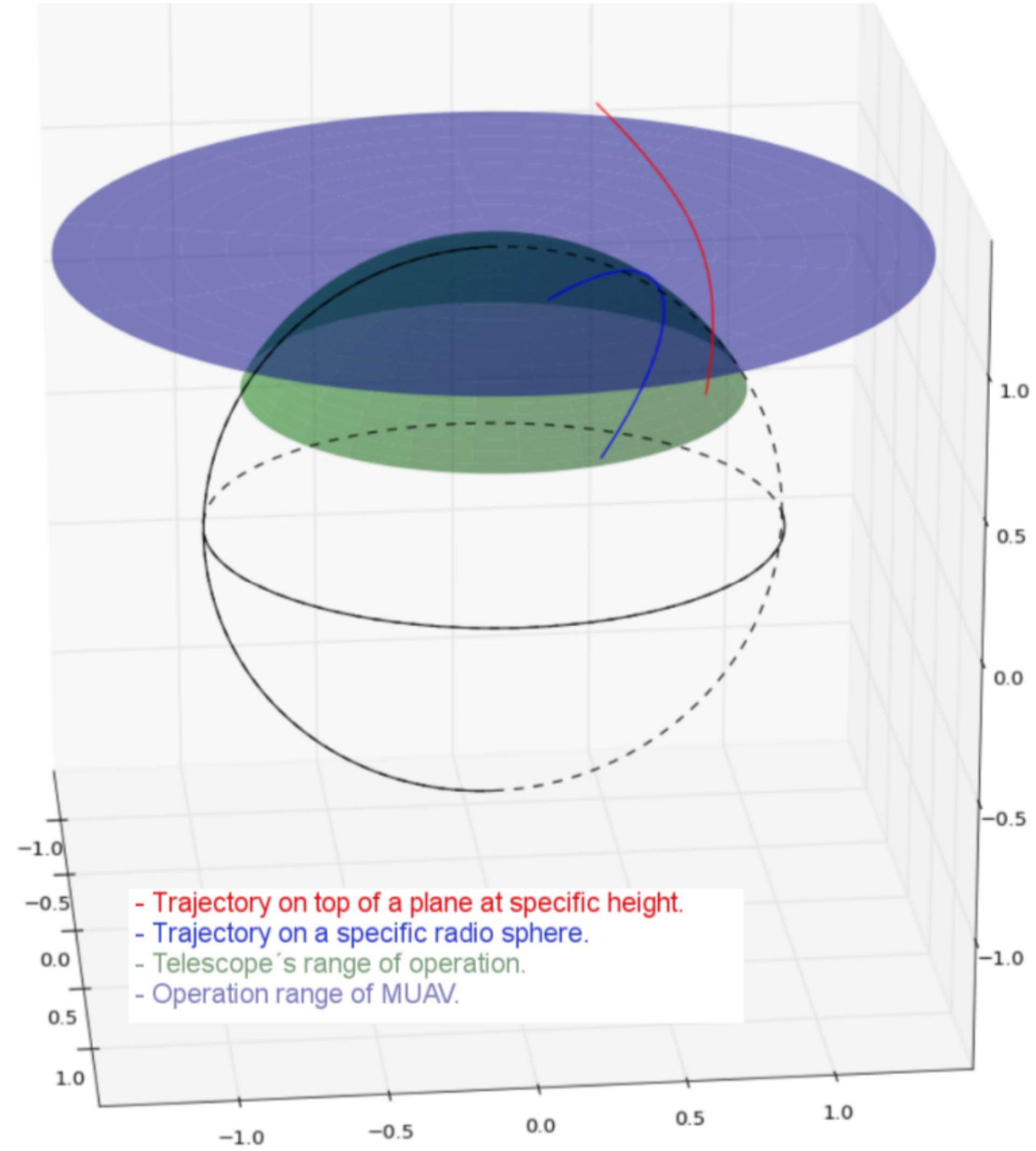}
\caption{Examples of trajectories for the MUAV.\label{fig:traj}}
\end{figure}

We considered positioning as the ability of the device to fly to a point in a three-dimensional space and stay in static flight with a specific orientation (hover). Also, we evaluated the potential of the device to follow a pre-established trajectory at a specific time. For this work, this trajectory corresponds to curved lines on planes with specific orientations, which will compute according to the position of the astronomical observatory on the Earth. Figure \ref{fig:traj} shows examples of the trajectories for the MUAV, the blue line represents the path on a sphere of constant radius, and the red line represents the path in a horizontal plane with constant height. Both paths match with the tracking of the same astronomical object. The typical operating range of astronomical telescopes, corresponding to a maximum angle range of $60^{\circ}$ from the zenith (surface marked in green), and the limitations inherent in the operating radius of the device (purple surface) must consider.

We consider the stability of the MUAV as the characteristic of the device to stay in a specific position or follow a given trajectory, with or without the presence of external disturbances (wind gusts). The RMS errors obtained when performing these maneuvers are considered as a quantification of the instability of the device. The latter is closely related to the implemented control techniques, the ability to obtain its position and orientation (POSE) within a time interval and the adequate response of its motors. The latest developments suggest that the improvements obtained in the flight and operational capabilities of the MUAVs depend on the sensors and devices that increase their capability to perceive the environment where they are located \citep{Kanellakis2017}.

\subsection{MUAV Flight Performance}

The flight operating requirements needed for the MUAV are: reach an initial position and perform a specific trajectory in a given time. This trajectory could be curvilinear one on a horizontal plane with a fixed height. Another possibility is a circular trajectory of a constant radius on a secant plane of the imaginary sphere (see, Figure \ref{fig:traj}). In addition, it is necessary to consider the presence of wind.

The first requirement is reported in \cite{Tran2015}. They tested a positioning control of a quadcopter and evaluated its stability while interacting with a wind of 5 m/s. The stability results were \textbf{0.41}, \textbf{0.19}, and \textbf{0.82 cm} for global coordinate axes X, Y, and Z, respectively.

For the trajectories in a horizontal plane, \cite{Razinkova2014} reports the stability of a quadcopter, which follows a path at a fixed height with the presence of constant and variable perturbations applied in different directions, resulting in stability of \textbf{14.77 cm} for \textbf{straight trajectories} and  \textbf{3.29 cm} for the \textbf{circular ones}.

For tilted trajectories with a constant radius, their inclination angle will depend on the apparent position of the astronomical object of interest. \cite{Jang2015} proved the capability of the quadcopters to follow this type of trajectories in specific times or specific speeds.  They reported maximum experimental errors of \textbf{13 cm} for a horizontal circular trajectory. However, their investigation does not include the performance of the device with external disturbances.

The above information offers us a theoretical overview (numerical simulations) of the MUAV capability to perform the required trajectories. However, since these devices are sub-acted systems\footnote{There are 4 inputs or controls (Pitch, Yaw, Roll, Thrust) for 6 responses or degrees of freedom.}, there always will be an error in the performing of these tasks. It is therefore that the implemented control techniques in the MUAVs play a fundamental role. 

Moreover, it is important to consider the errors produced by the resolution of the sensing techniques. An example of this is the positioning error of the MUAV when using global positioning systems (GPS, GLONASS, GALILEO, etc.). For these techniques, the uncertainty values are in the range of meters. However, in the last few years, techniques known as DGPS (Differential GPS) or RTK (Real Time Kinematics) have been developed, where the position determination for these devices in open environments (outdoor) is significantly improved, obtaining values below 5 cm \citep{Bing2015}. 

In the scientific literature, we can find experimental investigations where the techniques mentioned above are implemented, as in \cite{Joubert2016}, where trajectory tracking routines are programmed in a commercial quadcopter with an RTK system and whose results report errors of \textbf{1.1 cm} for external environments. Another example can be found in \cite{Lera2017}, in which differential systems are installed in a customized MUAV for the calibration of radio astronomical antennas, here the position of the device that carries a transmitting antenna is fundamental. The results obtained, experimentally, in the determination of the horizontal position show an average error values of \textbf{1 cm}.

Commercial MUAVs that use an RTK system have greater accuracy in their positioning, the specifications with these devices are \textbf{1 cm} for the \textbf{horizontal} position and \textbf{2 cm} for the \textbf{vertical} one within a one-kilometer radius (www.dji.com/d-rtk/info).

There is another research on the outdoor performance of a commercial MUAV without RTK system, where the \textbf{three-dimensional stability} of \textit{Matrice 100} in the presence of wind (3.6 -7.4 m/s) is reported for three specific routines: hover (\textbf{4.5 cm RMS}), step response (\textbf{26 cm RMS}) and path tracking (\textbf{10 cm RMS}) \citep{Sa2017}, which correspond to the routines of interest of this work. 

\begin{deluxetable}{c|c|c|c|c}[h!]
\tablecaption{MUAV fligh stability performance. \label{tabla:Perf}}
\tablecolumns{3}
\tablenum{5}
\tablewidth{0pt}
\tablehead{
\colhead{Data type} &
\colhead{Positioning} &
\colhead{Hover} &
\colhead{Horizontal} &
\colhead{Tilted} \\
\colhead{} &
\colhead{} &
\colhead{} &
\colhead{plane} &
\colhead{plane} 
}
\decimals
\startdata
\multirow{3}{*}{Simulated}    & X=0.41   &   & \multirow{2}{*}{S:14.7}   & \multirow{3}{*}{C:13} \\
                              &  Y=0.10   &  &\multirow{2}{*}{C:3.29}  &      \\
                              & Z= 0.82  &   &          &      \\
\hline              
\multirow{2}{*}{Experimental} & V: 1*  & \multirow{2}{*}{4.5}  & \multirow{2}{*}{C:1.1*}  & \multirow{2}{*}{S:10}        \\   
                              & H: 2*  &      &          &         \\   
\enddata

\tablenotetext{*}{Using RTK system.}
\tablecomments{\textbf{The values are in centimeters}. V: Vertical; \\H: Horizontal; C: Curved path; S: Straight path.}
\end{deluxetable}%

Table \ref{tabla:Perf} shows a summary of the stability results. In this table, we can see that a MUAV can perform the required trajectories. Besides, for the trajectories that are on a plane with constant height, can be done in the presence of wind. Similarly, we expected the total stability achieved by these devices is within the range of \textbf{centimeters} both for its positioning and for its displacement. We are sure that soon there will be commercial MUAVs with the capabilities to fulfill all the requirements to be a reference source for the GLAO system.

In the meantime, we propose to stabilize the light source on the MUAV switching between different LEDs in an array. The right position of light can be controlled by a quick camera mounted on an auxiliary telescope ($d<r_{0}$). This technique is like the used for the laser launch telescopes in AO systems.

\section{CONCLUSIONS}

In this work, we present the requirements to generate an artificial reference source for a GLAO system. Also, we mentioned the features of a Multirotor Unmanned Aerial Vehicle, which will carry this source.

Likewise, the restrictions for the reference source are established based on the characteristics of the atmospheric turbulence of the Ground Layer for a 6.5 m telescope at San Pedro Martir astronomical observatory in Mexico. It is also stipulated that the light provided by a standard 3 mm LED at 800 meters above the observatory, is enough to generate a light source with a controllable apparent magnitude lesser than $m_{v}= -0.23$. 

The performance of MUAV from scientific literature and of the commercial devices as well, show that these devices can perform the required trajectories in the presence of wind. The total stability achieved is in the range of centimeters both for its positioning and for its displacement. Nevertheless, it is necessary to use auxiliary systems to fully comply with the requirement that the MUAV keeps the light source within the field view of a wavefront sensor.


\end{document}